\documentclass[aip,jap,twocolumn,superscriptaddress,longbibliography,10pt]{revtex4-1}

\usepackage{graphicx}
\usepackage[usenames,dvipsnames,svgnames]{xcolor}
\usepackage{hyperref}
\hypersetup{
    pdfnewwindow=true,      
    colorlinks=true,        
    linkcolor=Blue,         
    citecolor=Blue,         
    filecolor=Blue,         
    urlcolor=Blue           
}

\begin{document}

\title{Lattice thermal conductivity of 16 elemental metals from molecular dynamics simulations with a unified neuroevolution potential}

\author{Shuo Cao}
\affiliation{Beijing Advanced Innovation Center for Materials Genome Engineering, University of Science and Technology Beijing, Beijing 100083, P. R. China}

\author{Ao Wang}
\email{ao.wang@uclouvain.be}
\affiliation{Global Institute of Future Technology, Shanghai Jiao Tong University, Shanghai 200240, P. R. China}
\affiliation{Institute of Condensed Matter and Nanosciences, Université catholique de Louvain, Chemin des Étoiles 8, 1348 Louvain-la-Neuve, Belgium}

\author{Zheyong Fan}
\email{brucenju@gmail.com} 
\affiliation{College of Physical Science and Technology, Bohai University, Jinzhou, P. R. China}

\author{Hua Bao}
\affiliation{Global Institute of Future Technology, Shanghai Jiao Tong University, Shanghai 200240, P. R. China}

\author{Ping Qian}
\affiliation{Beijing Advanced Innovation Center for Materials Genome Engineering,  University of Science and Technology Beijing, Beijing 100083, P. R. China}

\author{Ye Su}
\email{b2238966@ustb.edu.cn}
\affiliation{Beijing Advanced Innovation Center for Materials Genome Engineering,  University of Science and Technology Beijing, Beijing 100083, P. R. China}

\author{Yu Yan}
\affiliation{Beijing Advanced Innovation Center for Materials Genome Engineering,  University of Science and Technology Beijing, Beijing 100083, P. R. China}
\date{\today}

\begin{abstract}
Metals play a crucial role in heat management in electronic devices, such as integrated circuits, making it vital to understand heat transport in elementary metals and alloys.
In this work, we systematically study phonon thermal transport in 16 metals using the efficient homogeneous nonequilibrium molecular dynamics (HNEMD) method and the recently developed unified neuroevolution potential version 1 (UNEP-v1) for 16 metals and their alloys.
We compare our results with existing ones based on the Boltzmann transport equation (BTE) approach and find that our HNEMD results align well with BTE results obtained by considering phonon-phonon scattering only. 
By contrast, HNEMD results based on the conventional embedded-atom method potential show less satisfactory agreement with BTE ones.
Given the high accuracy of the UNEP-v1 model demonstrated in various metal alloys, we anticipate that the HNEMD method combined with the UNEP-v1 model will be a promising tool for exploring phonon thermal transport properties in complex systems such as high-entropy alloys.
\end{abstract}

\maketitle

\section{Introduction}

Metals are indispensable components in many electronic devices and thus play an important role in heat management.
Heat transport in metals is contributed by both electrons and phonons \cite{ashcroft1976book, ziman2001electrons}.
For most metals, electrons are the major heat carriers, but phonons are found to contribute a significant portion in some metals. 
For example, Tong \textit{et al.} investigated 11 metals and found that phonon thermal conductivity contributes between 1\% and 40\% to the total thermal conductivity \cite{TongZhen2019PRB}. 
Therefore, even though phonons are not the sole heat carriers in metals, it is still important to study phonon thermal transport in metals separately.

From the atomistic point of view, phonon thermal transport in metals have been mainly studied using two methods, one is based on Boltzmann transport equation (BTE) and anharmonic lattice dynamics, and the other based on molecular dynamics (MD) \cite{Gu2021jap}.
For the BTE method, the harmonic and anharmonic force constants are usually calculated using density functional theory (DFT) approaches. 
One of the advantages of the BTE method is its capability of including both phonon-phonon scattering and phonon-electron scattering. Another advantage of the BTE method is that electronic thermal transport can be treated at the same footing.
In view of these, phonon thermal transport in many metals has been studied using the BTE method \cite{Stojanovic2010PRB, WangYan2016JAP, Jain2016PRB, TongZhen2019PRB, Chen2019PRB, Wen2020PRB, Li2020PRB, Cui2021AMS, Zhang2022PRB, BaoHua2024APL}.

Despite of these advantages, the BTE method also has limitations.
The major limitation of the BTE method is its high order scaling of the computational cost with respect to the periodic calculation cell, restricting it to the study of relatively simple structures. 
On the other hand, the MD simulation approach has a linear scaling computational cost with respect to the periodic calculation cell and can be used to study realistically complex structures.
Moreover, MD simulation can naturally incorporate the full anharmonicity of phonon-phonon scattering, while the BTE approach is usually applied by considering three-phonon scattering only, and it entails considerably more efforts to consider four-phonon scattering \cite{feng2018prb, xia2018apl, Zhang2021jpcm, han2022cpc}.
One of the crucial inputs to MD methods is the interatomic potential, which is typically the embedded-atom method (EAM) potential \cite{Daw1984prb} for metals. 
While computationally efficient, the limitations of these empirical potentials have become increasingly apparent, leading to the emergence of alternatives based on machine-learning techniques such as neural network potentials \cite{Behler2007prl}.

\begin{figure*}[!]
\begin{center}
\includegraphics[width=1.8\columnwidth]{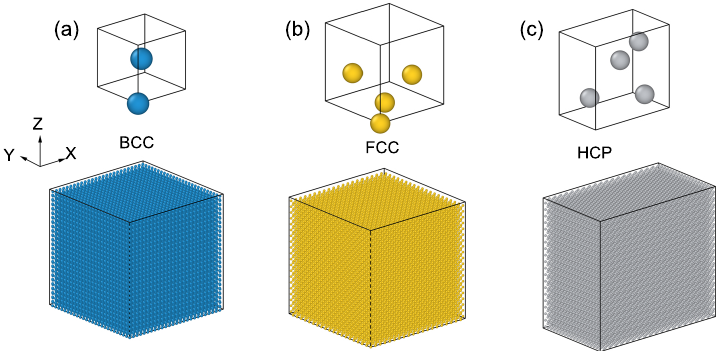}
\caption{Simulation models in this work. Unit cells and supercells for metals with (a) BCC,  (b) (FCC, and (c) HCP lattices. The supercells for the BCC and FCC lattices are cubic, while that for the HCP lattice is not cubic but orthogonal. The spacial directions $x$, $y$, and $z$ align with the cell axes of the lattices. The BCC supercell contains 35152 atoms, and the FCC and HCP supercells contain 32000 atoms.}
\label{figure:model}
\end{center}
\end{figure*}

Due to its high efficiency, the neuroevolution potential (NEP) approach \cite{fan2021prb} has found wide applications in heat transport \cite{Dong2024JAP}. 
However, these applications have thus far been limited to non-metal materials. 
Recently, Song \textit{et al.} \cite{song2024nc} developed a unified NEP model, UNEP-v1, for 16 metals and their alloys. 
The UNEP-v1 model has been shown to outperform traditional EAM potential \cite{zhou2004prb} in many physical properties, including elastic constants, surface energy, vacancy formation energy, melting point, and phonon dispersion. 
In this work, we utilize the UNEP-v1 model in MD simulations to systematically study heat transport in 16 metals, including five body-centered cubic (BCC) metals (Cr, Mo, Ta, V, W), eight face-centered cubic (FCC) metals (Ag, Al, Au, Cu, Ni, Pb, Pd, Pt), and three hexagonal close-packed (HCP) metals (Mg, Ti, Zr).
Phonon thermal conductivities of these metals are calculated using the efficient homogeneous nonequilibrium molecular dynamics (HNEMD) method \cite{Evans1982prl, fan2019prb} as implemented in the graphics processing units molecular dynamics (GPUMD) package \cite{FAN2017cpc}.
After examining the time convergence of the results and the effects of isotope scattering, we compare the MD results based on the UNEP-v1 model with the MD results based on the EAM potential as well as existing BTE results. 
It is found that the MD results based on the UNEP-v1 model align well with BTE results, while the MD results based on the EAM potential show less satisfactory agreement with BTE ones.
Given the high accuracy of the UNEP-v1 model demonstrated in various metal alloys, we anticipate that the HNEMD method combined with the UNEP-v1 model will be a promising tool for exploring heat transport properties in metal alloys.

\section{Models and Methods}

\subsection{Simulation models}

At ground state, five of the sixteen elemental metals we consider form BCC lattices (Cr, Mo, Ta, V, W), eight form FCC lattices (Ag, Al, Au, Cu, Ni, Pb, Pd, Pt), and three form HCP lattices (Mg, Ti, Zr). 
Figure~\ref{figure:model} illustrates the orthogonal unit cells for each type of lattice and the corresponding supercells used in the HNEMD simulations. 
For BCC lattices, the supercells consist of $26 \times 26 \times 26$ unit cells, totaling 35152 atoms. For FCC and HCP lattices, the supercells consist of $20 \times 20 \times 20$ unit cells, totaling 32000 atoms.

\subsection{Interatomic potentials}

We will mainly use the NEP model \cite{fan2021prb} in the MD simulations, but will also use the EAM potential \cite{zhou2004prb} for comparison.

\subsubsection{The embedded-atom method potential}

For the EAM potential, we use the widely used parameterization by Zhou \textit{et al.} \cite{zhou2004prb} for all the 16 elements considered in this work.
In this potential, the potential energy of atom $i$ is expressed as
\begin{equation}
U_i = \frac{1}{2} \sum_{j\neq i} \phi(r_{ij}) + F (\rho_i).
\end{equation}
Here, $\phi(r_{ij})$ represents a pairwise potential as a function of the distance $r_{ij}$ between atoms $i$ and $j$, and $F(\rho_i)$ is the so-called embedding energy, which is a functional of the electron density function
\begin{equation}
\rho_i = \sum_{j\neq i} f(r _{ij}).
\end{equation}
Explicit expressions for the functions $\phi$, $F$, and $f$ can be found in the original work \cite{zhou2004prb}. 
This version of EAM potential has been implemented in the GPUMD package \cite{FAN2017cpc} used in the present work.

\subsubsection{The neuroevolution potential}

The NEP approach \cite{fan2021prb,Fan2022jpcm,fan2022jcp,song2024nc} is a highly efficient machine-learned potential as implemented in the GPUMD package \cite{FAN2017cpc}.
In this method, the potential energy of atom $i$ can be expressed as
\begin{equation}
U_i = U_i( \mathbf{q}^i),
\end{equation}
where $\mathbf{q}^i$ is the descriptor vector for atom $i$. 
NEP is a neural network potential model with a single hidden layer, and the site potential energy can be explicitly written as:
\begin{equation}
U_i = \sum_{\mu=1} ^{N_\mathrm{neu}}w ^{(1)} _{\mu}  \tanh\left(\sum_{\nu=1} ^{N_\mathrm{des}} w ^{(0)}_{\mu\nu} q^i_{\nu} - b^{(0)}_{\mu}\right) - b^{(1)}.
\end{equation}
Here, $N_\mathrm{des}$ is the dimension of the descriptor, $N_\mathrm{neu}$ is the dimension of the hidden layer, $\tanh$ is the activation function for the hidden layer neurons, and $\mathbf{w}^{(0)}$, $\mathbf{w}^{(1)}$, $\mathbf{b}^{(0)}$, and $b^{(1)}$ are the trainable weight and bias parameters in the neural network.
The descriptor used in NEP has been detailed previously \cite{fan2022jcp}.
The descriptor also contains a number of trainable parameters \cite{Fan2022jpcm,fan2022jcp} that are optimized during the training process.
The training of NEP is driven by a loss function calculated based on predicted and target values for energy, force, and virial stress. 

In this work, we use the UNEP-v1 model developed by Song 
 \textit{et al.} \cite{song2024nc}.
Although we will focus on the elemental elements, we note that this UNEP-v1 model applies to 16 elemental metals (Ag, Al, Au, Cr, Cu, Mg, Mo, Ni, Pb, Pd, Pt, Ta, Ti, V, W, Zr) and their arbitrary alloys.
It has been shown that the UNEP-v1 model outperforms the EAM potential by Zhou \textit{et al.} \cite{zhou2004prb} in terms of many physical properties, including elastic constant, surface formation energy, vacancy formation energy, melting point, and phonon dispersion \cite{song2024nc}. 
It is one of the purposes of this work to further explore the capabilities of the UNEP-v1 model in phonon thermal transport properties.  

\subsection{Heat transport methods within molecular dynamics}

\subsubsection{Heat current}

Both the EAM and NEP models are many-body interatomic potentials.
For both potentials, the total potential energy is the sum of the site potential energies:
\begin{equation}
U = \sum_i U_i.
\end{equation}
The force acting on atom $i$ can be derived to be \cite{fan2015prb}
\begin{equation}
\mathbf{F}_i=\sum_{j\neq i}\left(\frac{\partial U_i}{\partial \mathbf{r}_{ij}}-\frac{\partial U_j}{\partial \mathbf{r}_{ji}}\right),
\end{equation}
where $\mathbf{r}_{ij} = \mathbf{r}_j - \mathbf{r}_i$, $\mathbf{r}_i$ being the position of atom $i$.
Based on this force expression, the heat current related to interactions has been derived \cite{fan2015prb} to be
\begin{equation}
    \mathbf{J} = \sum_i \mathbf{W}_i \cdot \mathbf{v}_i,
\end{equation}
where $\mathbf{v}_i$ is the velocity of atom $i$ and 
\begin{equation}
    \mathbf{W}_i = \sum_{j\neq i} \mathbf{r}_{ij} \otimes \frac{\partial U_j} {\partial \mathbf{r}_{ji}}
\end{equation}
is the per-atom virial for atom $i$.

\subsubsection{The homogeneous non-equilibrium molecular dynamics method}

With the availability of the heat current, we can calculate the phonon thermal conductivity using various MD methods.
 In this work, we will mainly use the HNEMD method \cite{fan2019prb}, which has been shown to be very efficient and has become one of the canonical methods for computing the phonon thermal conductivity in the diffusive transport regime \cite{Gu2021jap}.
In this method, a net heat flow is generated by applying an external driving force:
\begin{equation}
\mathbf{F}_i^{\rm ext}= \mathbf{F}_{\rm e} \cdot \mathbf{W}_i,
\end{equation}
where $F_{\rm e}$ is a parameter of the dimension of inverse length. 
The ensemble average of the induced heat current is proportional to the driving force parameter, 
\begin{equation}
\label{equation:kappa_hnemd}
    \langle J^{\alpha} \rangle = TV \sum_{\beta} \kappa^{\alpha\beta} F_{\rm e}^{\beta},
\end{equation}
where $T$ is the temperature and $V$ is the volume of the system.
Base on this relation, the thermal conductivity tensor $\kappa^{\alpha\beta}$ can be computed. 

\subsection{The spectral thermal conductivity}

In the framework of the HNEMD method, one can calculate spectral thermal conductivity using the virial-velocity correlation function \cite{fan2019prb}
\begin{equation}
    \mathbf{K}(t) = \left\langle \sum_i \mathbf{W}_i(0) \cdot \mathbf{v}_i(t) \right\rangle.
\end{equation}
The spectral thermal conductivity $\kappa^{\alpha\beta}(\omega)$ as a function of the phonon frequency $\omega$ can be calculated from the following relation:
\begin{equation}
\label{equation:kappa_omega}
\frac{2}{VT} \int_{-\infty}^{+\infty} e^{i\omega t} K^{\alpha}(t) \text{d}t= \sum_{\beta} \kappa^{\alpha\beta}(\omega){F_{\rm e}}^\beta.
\end{equation}

\begin{table*}[!]
\centering \setlength{\tabcolsep}{0.9mm}\renewcommand{\arraystretch}{1.7} 
\caption{The phonon thermal conductivities of 16 metals and 4 binary alloys calculated using various methods, including BTE considering phonon-phonon (pp) scattering only, BTE considering both pp and phonon-electron (pe) scattering, HNEMD with the UNEP-v1 model without considering isotope disorder (denoted as NEP), HNEMD with the UNEP-v1 model considering isotope disorder (denoted as NEP-isotope), HNEMD with the EAM potential without considering isotope disorder (denoted as EAM), and HNEMD with the EAM potential considering isotope disorder (denoted as EAM-isotope).
The numbers within the parentheses represent the uncertainty in the last digit.}
\begin{tabular}{ccccccccc}
\hline
\hline
&Element  &BTE (pp) &BTE (pp-pe) &NEP &NEP-isotope &EAM &EAM-isotope \\
\hline
&Ag    
&6.03\cite{TongZhen2019PRB},5.2\cite{WangYan2016JAP},3.75\cite{Jain2016PRB}   &5.69\cite{TongZhen2019PRB},5.2\cite{WangYan2016JAP},4\cite{Jain2016PRB},4.5\cite{BaoHua2024APL},9.3\cite{Stojanovic2010PRB}
&3.4(1)  &3.5(1) &4.2(1) &4.2(1)\\
&Al  
&10.02\cite{TongZhen2019PRB},5.8\cite{WangYan2016JAP}8.19\cite{Jain2016PRB}  &8.95\cite{TongZhen2019PRB},5.8\cite{WangYan2016JAP},6\cite{Jain2016PRB},21.1\cite{Stojanovic2010PRB},20.22\cite{Cui2021AMS},9.19\cite{Zhang2022PRB},10.67\cite{Li2020PRB} 
&7.2(1)  &- &2.6(1) &-\\
&Au   
&3.05\cite{TongZhen2019PRB},2.6\cite{WangYan2016JAP},1.53\cite{Jain2016PRB}   &2.80\cite{TongZhen2019PRB},2.6\cite{WangYan2016JAP},2\cite{Jain2016PRB},5\cite{Stojanovic2010PRB} & 2.0(1) &- &2.9(1)  &- \\
&Cu  
&19.49\cite{TongZhen2019PRB},16.9\cite{WangYan2016JAP},11.78\cite{Li2020PRB} &17.42\cite{TongZhen2019PRB},16.9\cite{WangYan2016JAP},11.6\cite{BaoHua2024APL},22.2\cite{Stojanovic2010PRB},11.78\cite{Li2020PRB}
&13(1)  &13(1) & 6.9(2)  &6.9(1)\\
&Ni   &27.79\cite{TongZhen2019PRB},33.2\cite{WangYan2016JAP} &15.33\cite{TongZhen2019PRB},23.2\cite{WangYan2016JAP},9.6\cite{Stojanovic2010PRB} 
&30(1)  &27(1) &27(1)  &25(1)\\
&Pb   
&    & & 0.6(1)  &0.6(1) &0.6(1)  &0.5(1)\\ 
&Pd  
&19.62\cite{TongZhen2019PRB}  &12.51\cite{TongZhen2019PRB} 
& 11(1)  &11(1) & 8.3(1)  &8.1(1)  \\
&Pt  
&8.67\cite{TongZhen2019PRB},7.1\cite{WangYan2016JAP}  &6.49\cite{TongZhen2019PRB},5.8\cite{WangYan2016JAP},8.3\cite{Stojanovic2010PRB},3.72\cite{Zhang2022PRB} 
&7.7(2) &8.0(1)   &7.3(1) &7.2(1) \\
&Mo\ 
&84.17\cite{BaoHua2024APL},110.85\cite{Wen2020PRB} 
&26\cite{BaoHua2024APL} ,37.01\cite{Wen2020PRB}   
&182(3)  &75(3) &90(7)   &63(5)\\
&Cr 
&     &  & 207(7)   &166(3) & 75(3)&66(1) \\
&Ta  &    &   & 59(2)  &- &  24(1)&-\\
&V  &      & &  97(3)  &- &9.4(2)&-\\ 
&W   
& 123\cite{BaoHua2024APL},218.21\cite{Chen2019PRB}
&44.8\cite{BaoHua2024APL},42.2\cite{Stojanovic2010PRB},63.37\cite{Zhang2022PRB},46\cite{Chen2019PRB}    
&164(7)  &115(4) &  110(3) &99(4) \\
&Ti-x 
&11.92\cite{TongZhen2019PRB} 
&5.32\cite{TongZhen2019PRB}  
& 15(1) &14(1) & 10(1) &9.7(1)\\
&Ti-z    
&11.92\cite{TongZhen2019PRB} 
&5.32\cite{TongZhen2019PRB}  
& 19(1)  &18(1)& 11(1) &11(1)\\
&Zr-x   
&   &    &  8.3(1)   &8.2(1)  &6.4(1) & 6.4(1)\\
&Zr-z  
&   &    &  8.9(1) &8.5(1) &7.9(1) &7.5(1)\\
&Mg-x   
&9.27\cite{TongZhen2019PRB} 
&7.15\cite{TongZhen2019PRB},8.56\cite{Cui2021AMS}  
& 5.6(1) &5.2(1) &  2.6(1)  &2.5(1)\\
&Mg-z 
&9.27\cite{TongZhen2019PRB} 
&7.15\cite{TongZhen2019PRB},8.56\cite{Cui2021AMS}  
& 5.7(1) &5.4(1) &  2.7(1) &2.7(1) \\
&Cu3Au 
&3.32\cite{TongZhen2019PRB} &1.89\cite{TongZhen2019PRB} 
&2.9(1) &- &  2.6(1) &- \\
&CuAu 
&3.32\cite{TongZhen2019PRB} &2.32\cite{TongZhen2019PRB} 
&2.5(1) &- & 3.1(1) &- \\
&Ni3Al
&7.78\cite{TongZhen2019PRB} &4.72\cite{TongZhen2019PRB} 
&13(1) &- & 5.0(1) &- \\
&NiAl 
&12.31\cite{TongZhen2019PRB} &6.02\cite{TongZhen2019PRB} 
&7.2(1) &- &  1.2(1) &- \\
\hline
\hline
\end{tabular}
\label{table:kappa}
\end{table*}

\subsection{Details of molecular dynamics simulations}

All the MD simulations were performed using the GPUMD software \cite{FAN2017cpc}. 
In the HNEMD simulations, all the three directions of a supercell adopt the periodic boundary condition.
For time integration, we used a time step of 4 fs, which was tested to be sufficiently small for all the metals.
Before conducting a HNEMD simulation, the system was first equilibrated in the isothermal-isobaric ensemble for 1 ns to achieve the target temperature and pressure. 
In this work, we considered the zero pressure condition with varying temperatures.
Then during the HNEMD simulation, the isothermal-isochoric ensemble was used.
The production time in each HNEMD simulation was 10 ns.
The driving force parameter $F_{\rm e}$ in the HNEMD simulations muse be sufficiently small to maintain the system within the linear response regime, and large enough to ensure a high signal-to-noise ratio.
With extensive tests, we have determined appropriate values of it.
Specifically, we have found that a value of $F_{\rm e}=1$ $\mu$m$^{-1}$ is appropriate for both FCC and HCP metals, while a smaller value of $F_{\rm e}=0.1$ $\mu$m$^{-1}$ is more adequate for BCC metals with significantly higher phonon thermal conductivity.
For BCC and FCC metals, the phonon thermal transport is isotropic and we have thus calculated a single component.
For HCP metals, there are generally two independent components that were separated calculated.

\begin{table*}[!]
\centering\setlength{\tabcolsep}{1.2mm}\renewcommand{\arraystretch}{1.5}
\caption{Natural isotope abundances for the 16 metals considered in this work \cite{Haynes2016book}.}
\begin{tabular}{c|cc|cc|cc|cc|cc|cc|cc}
\toprule
Element&\multicolumn{14}{c}{Abundance (\%)}\\
\hline                                                  
Ag &$^{107}$Ag &51.8    &$^{109}$Ag &48.2   &           &         &           &         &           &        &          &         &           &\\
Al &$^{27}$Al  &100.0   &           &         &           &         &           &         &           &        &          &         &           &\\
Au &$^{197}$Au &100.0   &           &         &           &         &           &         &           &        &          &         &           &\\
Cu &$^{63}$Cu  &69.2    &$^{65}$Cu  &30.9   &           &         &           &         &           &        &          &         &           &\\
Ni &$^{58}$Ni  &68.1    &$^{60}$Ni  &26.2   &$^{61}$Ni  &1.1    &$^{62}$Ni  &3.6    &$^{64}$Ni  &0.9   &          &         &           &\\
Pb &$^{204}$Pb &1.4     &$^{206}$Pb &24.1   &$^{207}$Pb &22.1   &$^{208}$Pb &52.4   &           &        &          &         &           &\\
Pt &$^{192}$Pt &0.8     &$^{194}$Pt &32.9   &$^{195}$Pt &33.8   &$^{196}$Pt &25.2   &$^{198}$Pt &7.4   &          &         &           &\\
Mo &$^{92}$Mo  &14.7    &$^{94}$Mo  &9.2    &$^{95}$Mo  &15.9   &$^{96}$Mo  &16.7   &$^{97}$Mo  &9.6   &$^{98}$Mo &24.3   &$^{100}$Mo &9.7\\
Cr &$^{50}$Cr  &4.3     &$^{52}$Cr  &83.8   &$^{53}$Cr  &9.5    &$^{54}$Cr  &2.4    &           &        &          &         &           &\\
Ta &$^{182}$Ta &100.0   &           &         &           &         &           &         &           &        &          &         &           &\\
V  &$^{50}$V   &0.3      &$^{51}$V   &99.8   &           &         &           &         &           &        &          &         &           &\\
W  &$^{180}$W  &0.1     &$^{182}$W  &26.5   &$^{183}$W  &14.3   &$^{184}$W  &30.6   &$^{186}$W  &28.4  &          &         &           &\\
Ti &$^{46}$Ti  &8.3     &$^{47}$Ti  &7.4    &$^{48}$Ti  &73.7   &$^{49}$Ti  &5.4    &$^{50}$Ti  &5.2   &          &         &           &\\
Zr &$^{90}$Zr  &51.5    &$^{91}$Zr  &11.2   &$^{92}$Zr  &17.1   &$^{94}$Zr  &17.4   &$^{96}$Zr  &2.8   &          &         &           &\\
Mg &$^{24}$Mg  &79.0    &$^{25}$Mg  &10.0   &$^{26}$Mg  &11.0   &           &         &           &        &          &         &           &\\
\hline
\end{tabular}
\label{table:abundance}
\end{table*}

\section{Results and Discussion}

\subsection{Time convergence of the phonon thermal conductivity}

\begin{figure}[htb]
\begin{center}
\includegraphics[width=0.9\columnwidth]{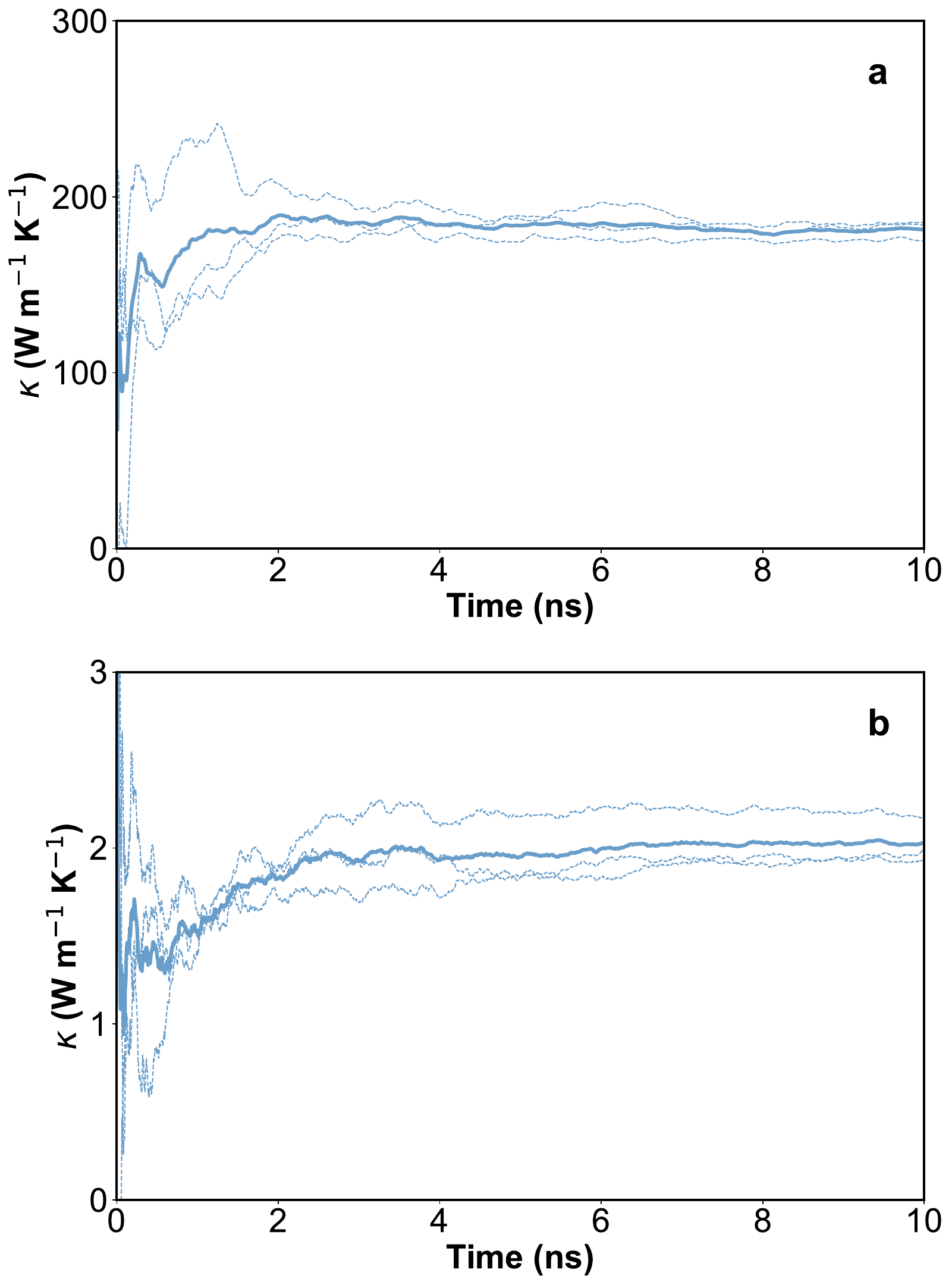}
\caption{Time convergence of the phonon thermal conductivity.
Cumulative averages of the phonon thermal conductivity $\kappa$ of (a) BCC Mo and (b) FCC Au as a function of time calculated using the UNEP-v1 model at 300 K and zero pressure. In both panels, the thin lines represent results from three independent HNEMD simulations, and the thick line represents the average of them. Isotope disorder was not considered in the simulations. }
\label{figure:kappa_t}
\end{center}
\end{figure}

We have systematically calculated the phonon thermal conductivity values for the 16 elemental metals using HNEMD simulations \cite{fan2019prb}.
While the running phonon thermal conductivity can be calculated from Eq.~(\ref{equation:kappa_hnemd}), it is beneficial to consider its cumulative average.
For phonon thermal transport in a given direction, the cumulative average can be calculated as 
\begin{equation}
    \kappa(t) = \frac{1}{t} \int_0^t \frac{\langle J(\tau)\rangle}{TVF_{\rm e}} \text{d} \tau.
\end{equation}
Here $J$ and $F_{\rm e}$ should be understood as the the components of the heat current vector and the driving force parameter vector, respectively, in the considered direction. 
Figure~\ref{figure:kappa_t} shows the $\kappa(t)$ values for two typical metals, BCC Mo and (b) FCC Au, calculated using the UNEP-v1 model at 300 K and zero pressure.
For both materials, $\kappa(t)$ converges well with respect to $t$ up to 10 ns, without a sign of divergence, which means that the systems are within the linear-response regime.
Notably, the converged phonon thermal conductivity of BCC Mo is about two orders of magnitude larger than that of FCC Au.
Based on the three independent runs, we can estimate an average phonon thermal conductivity and a statistical error (taken as the standard error) for each metal.
For BCC Mo, it is $182 \pm 3$ W m$^{-1}$ K$^{-1}$; for FCC Au, it is $2.02 \pm 0.08$ W m$^{-1}$ K$^{-1}$.
This large difference in the phonon thermal conductivity is partially reflected by the fact that the $F_{\rm e}$ value for BCC Mo is ten times smaller than  that for FCC Au, which means that the typical phonon mean free paths in BCC Mo are much larger than those in FCC Au.
The calculated phonon thermal conductivities for all the 16 metals are listed in Table~\ref{table:kappa}.

\begin{figure}[htb]
\begin{center}
\includegraphics[width=\columnwidth]{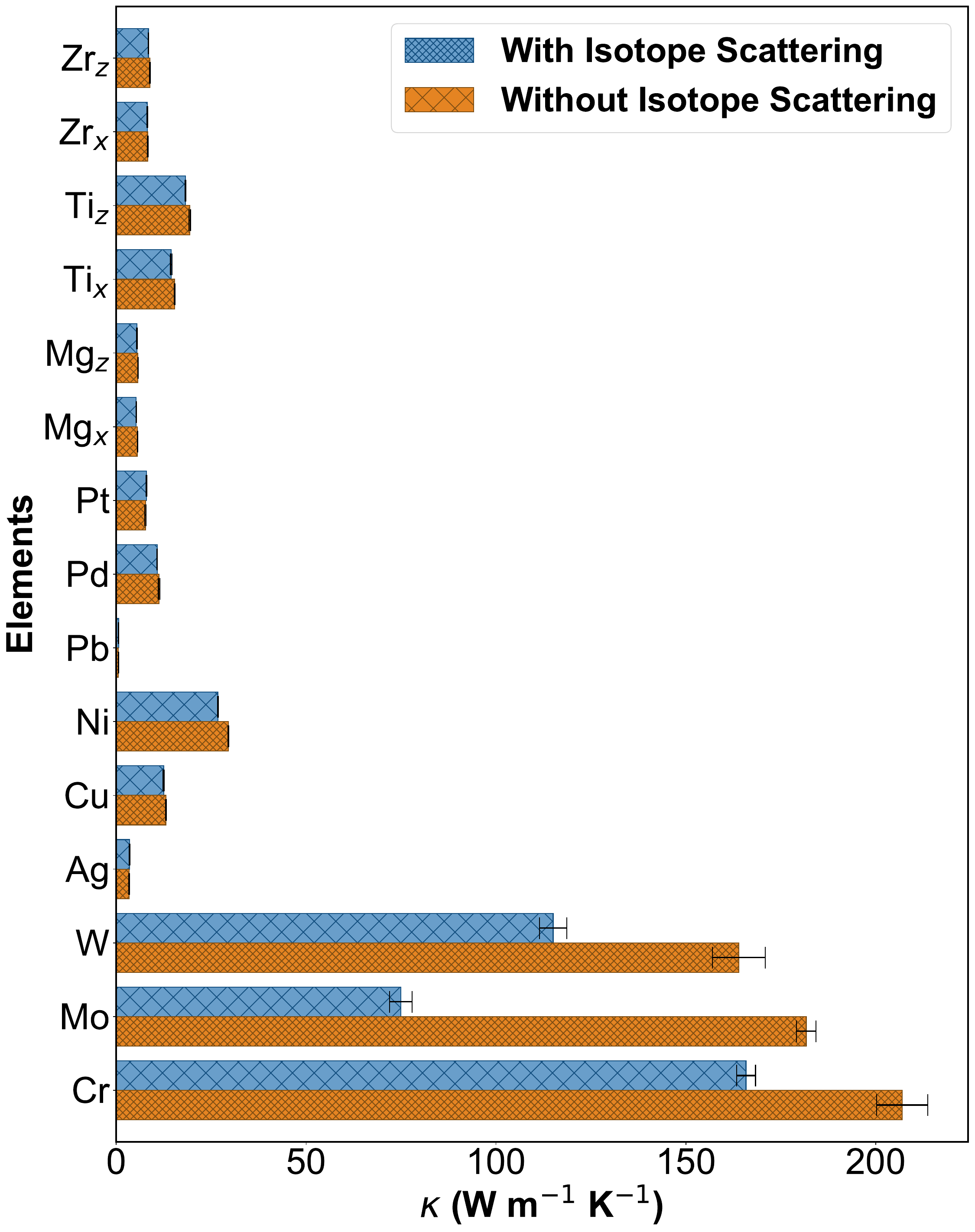}
\caption{Comparison of phonon thermal conductivities with and without isotope scattering effects. In all the MD simulations, the temperature is 300 K and the pressure is zero. Error bars denote the standard error of the mean from three independent calculations.}
\label{figure:isotope}
\end{center}
\end{figure}

\begin{figure}[htb]
\begin{center}
\includegraphics[width=1\columnwidth]{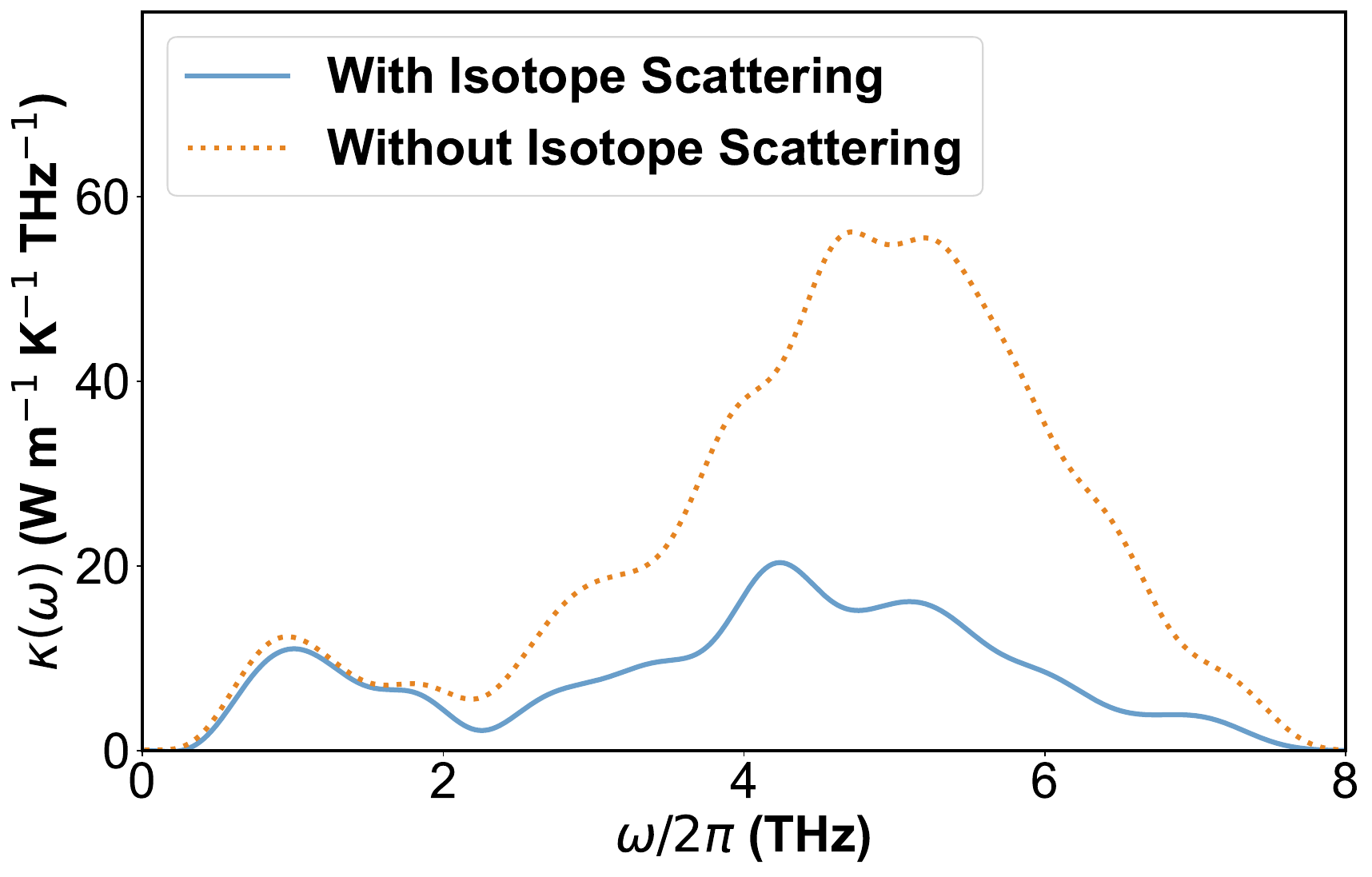}
\caption{Spectral thermal conductivity $\kappa(\omega)$ of Mo as a function of the phonon frequency $\omega/2\pi$.
The solid and dashed lines represent results obtained with and without considering isotope scattering, respectively. The simulation temperature is 300 K and the pressure is zero.} 
\label{figure:spectral}
\end{center}
\end{figure}

\subsection{Influence of isotope disorder on the phonon thermal conductivity}

In the above calculations, we have not considered isotope disorder.
For the 16 metals we considered, many of them have a few different isotopes with abundance over $0.3\%$, except for Al, Au, Ta and V.
It is therefore important to consider isotope disorder in the MD simulations. 
To this end, we assigned masses to the atoms in the MD simulations according to the natural abundances of the isotopes \cite{Haynes2016book} (see Table~\ref{table:abundance}).
The calculated phonon thermal conductivities for all the 16 metals considering isotope disorder are listed in Table~\ref{table:kappa}.

Figure~\ref{figure:isotope} compares the phonon thermal conductivities calculated based on HNEMD simulations with and without considering isotope disorder, ignoring the four metals (Al, Au, Ta and V) mentioned above.
It is clear that isotope disorder only has strong effects on the three BCC metals, namely, Cr, Mo, and W.
The BCC metal Mo is the most affected, with the phonon thermal conductivity reduced by a factor of $62\%$, from $182 \pm 3$ W m$^{-1}$ K$^{-1}$ without considering isotope disorder to $69 \pm 4$ W m$^{-1}$ K$^{-1}$ considering isotope disorder.
This strong reduction of phonon thermal conductivity has two reasons.
The first reason is that Mo has 7 isotopes, ranging from $^{92}$Mo to $^{100}$Mo, each with a considerable abundance.
The second reason is that isotopically pure Mo has large phonon thermal conductivity and thus weak phonon anharmonicity, which means that isotope disorder has strong effects in this material. 
To gain more insights on the significant effect of isotope disorder, we examine the spectral phonon thermal conductivity as presented in Fig.~\ref{figure:spectral}.
It shows that for isotopically pure Mo, the phonon thermal conductivity is mainly contributed by phonons with frequencies larger than $\nu=2$ THz.
Because isotope disorder induced phonon scattering rate is proportional to the square of phonon frequency \cite{Tamura1983prb}, isotope disorder has large effects on the phonon thermal conductivity of Mo.
The large effects of isotope scattering in Mo have also been verified based on the BTE approach \cite{Wen2020PRB}.

\subsection{Comparison between NEP-MD and EAM-MD results}

\begin{figure}[htb]
\begin{center}
\includegraphics[width=\columnwidth]{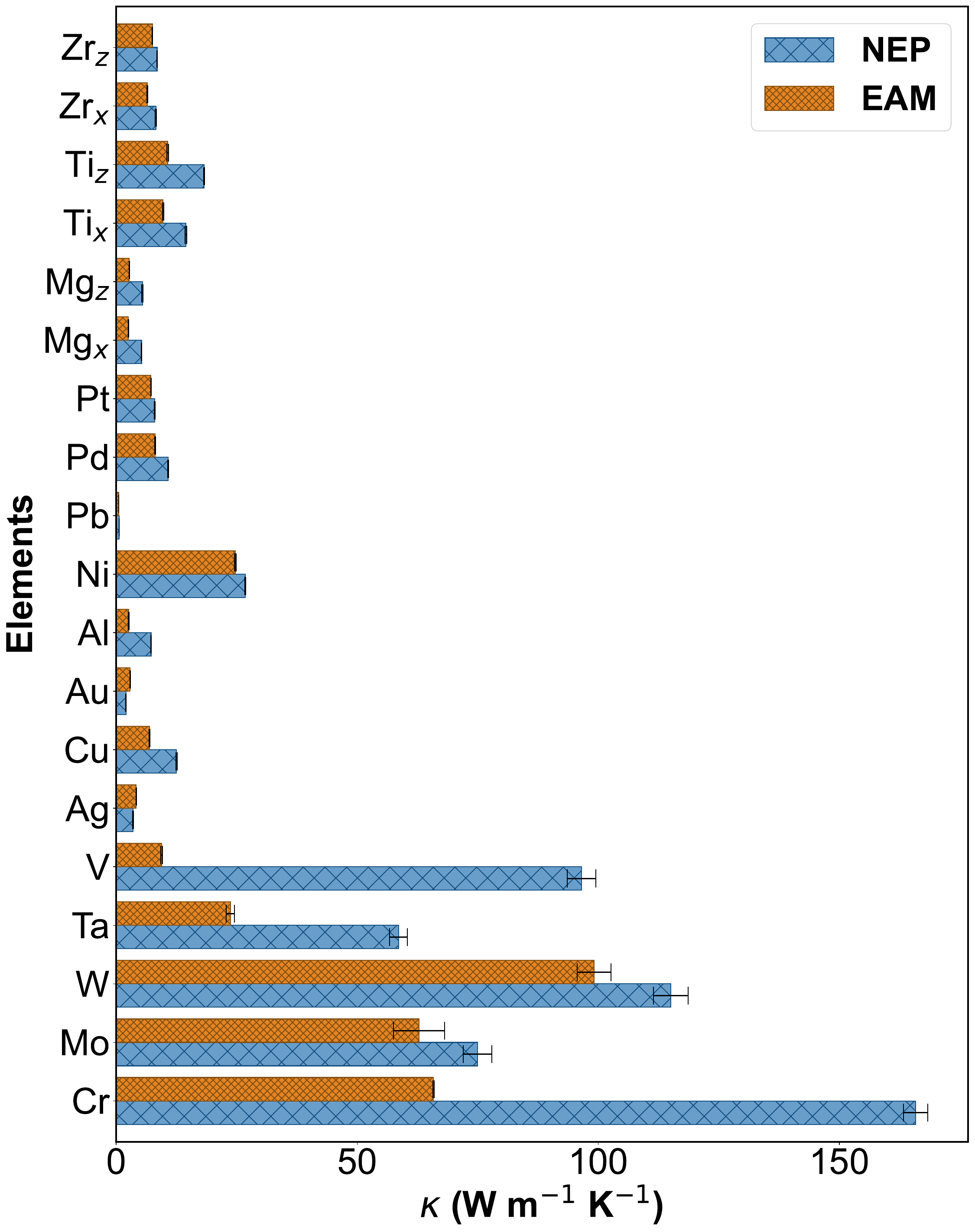}
\caption{Phonon thermal conductivity comparison between MD simulations with UNEP-v1 and EAM models.
In all the MD simulations, the temperature is 300 K, the pressure is zero, and isotope scattering is considered. Error bars denote the standard error of the mean from three independent calculations.}
\label{figure:nep_vs_eam}
\end{center}
\end{figure}

\begin{figure*}[htb]
\begin{center}
\includegraphics[width=2\columnwidth]{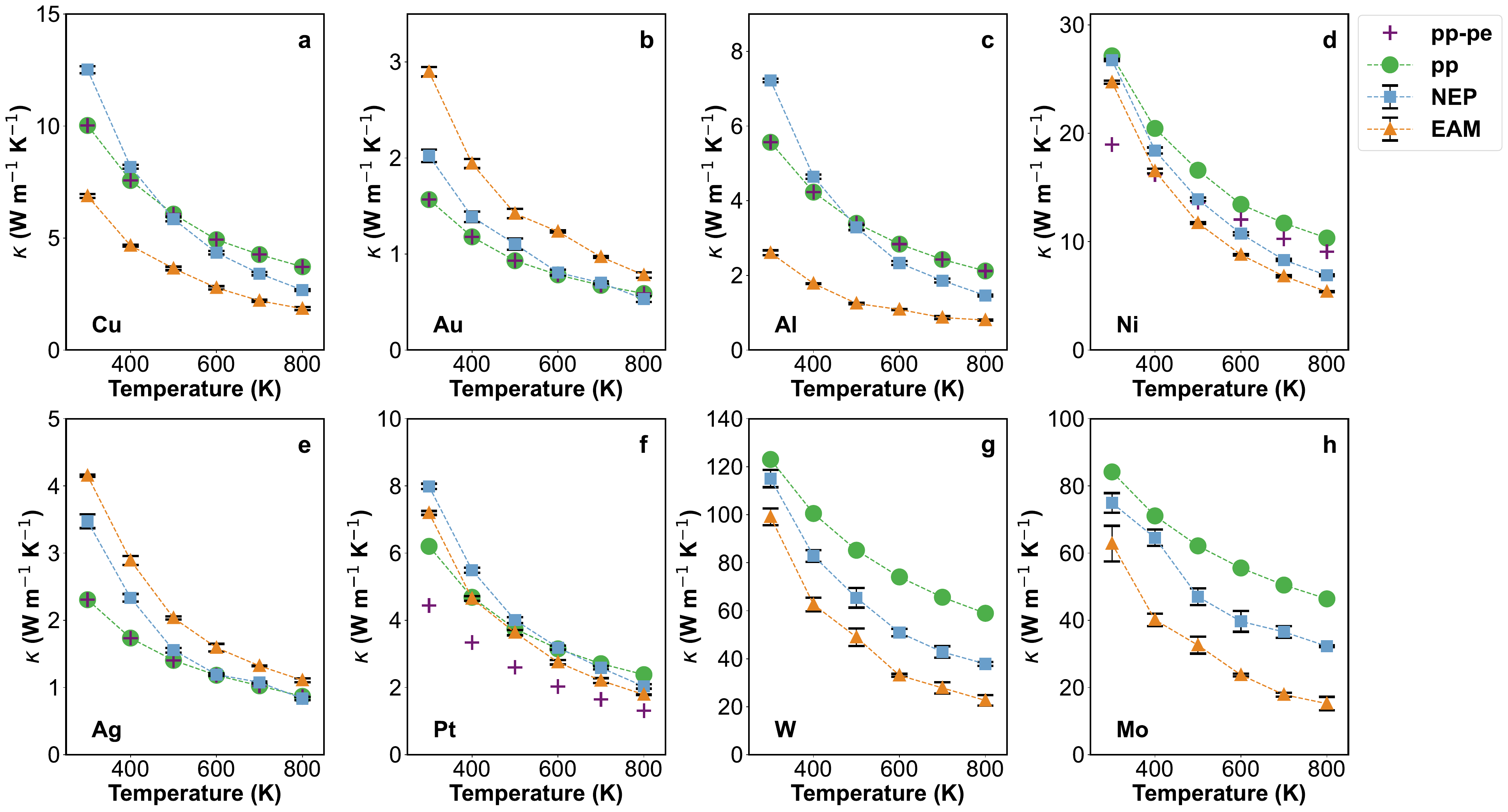}
\caption{Phonon thermal conductivity comparison between MD simulations and BTE calculations. In each panel, filled circles represent BTE results considering phonon-phonon (pp) scattering only, plus symbols represent BTE results considering both phonon-phonon and phonon-electron (pe) scatterings, filled squares represent MD results based on the UNEP-v1 model, and filled triangles represent MD results based on the EAM potential. 
BTE results for W and Mo are from Wang and Bao \cite{BaoHua2024APL}, while those for the remaining elements are from Wang \textit{et al.} \cite{WangYan2016JAP}.}
\label{figure:temperature}
\end{center}
\end{figure*}

After examining the effects of isotope scattering, we next compare the results obtained from the UNEP-v1 model \cite{song2024nc} with those from the empirical EAM potential \cite{zhou2004prb}, both considering isotope scattering in the MD simulations.
The results are presented in Fig.~\ref{figure:nep_vs_eam}.
Generally, the calculated phonon thermal conductivities from the EAM potential are smaller than those from the NEP model, particularly for the BCC metals. 
For V, the phonon thermal conductivities from the two potentials differ by one order of magnitude.
In view of the fact that the UNEP-v1 model by Song \textit{et al.} \cite{song2024nc} outperforms the EAM potential by Zhou \textit{et al.} \cite{zhou2004prb} in many physical properties, including the phonon dispersions, we believe that the phonon thermal conductivity results obtained by the UNEP-v1 model are generally more reliable. 
To gain deeper understanding on this problem, we next compare our MD results to existing BTE ones. 

\begin{figure}[htb]
\begin{center}
\includegraphics[width=0.9\columnwidth]{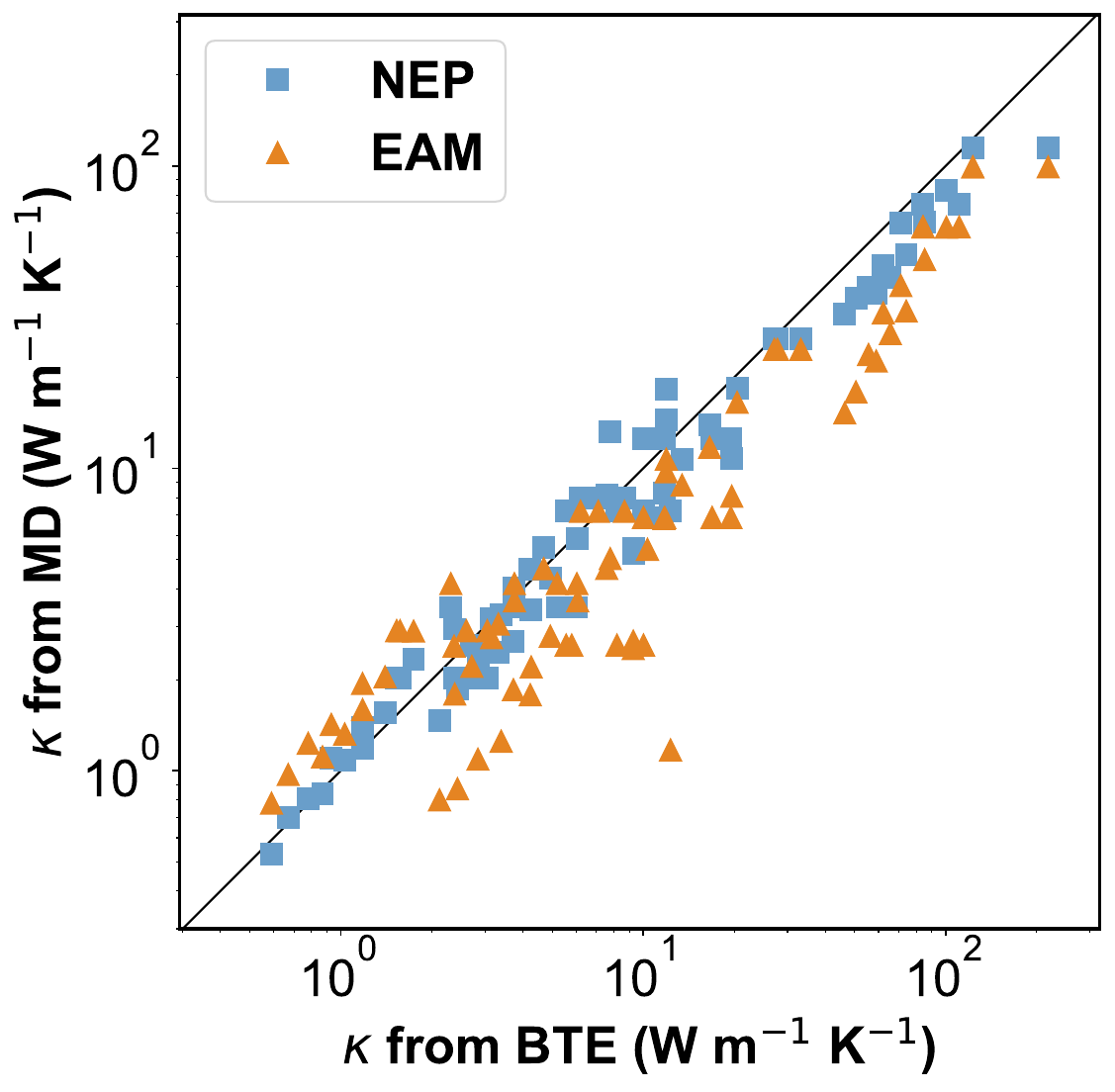}
\caption{Comparison of MD and BTE results for phonon thermal conductivities for all the materials considered in this work, including 16 metals and 4 binary alloys. Filled squares and triangles represent MD results from the UNEP-v1 model and the EAM potential, respectively. Only phonon-phonon scattering is considered in the BTE results.}
\label{figure:parity}
\end{center}
\end{figure}

\begin{figure}[htb]
\begin{center}
\includegraphics[width=0.9\columnwidth]{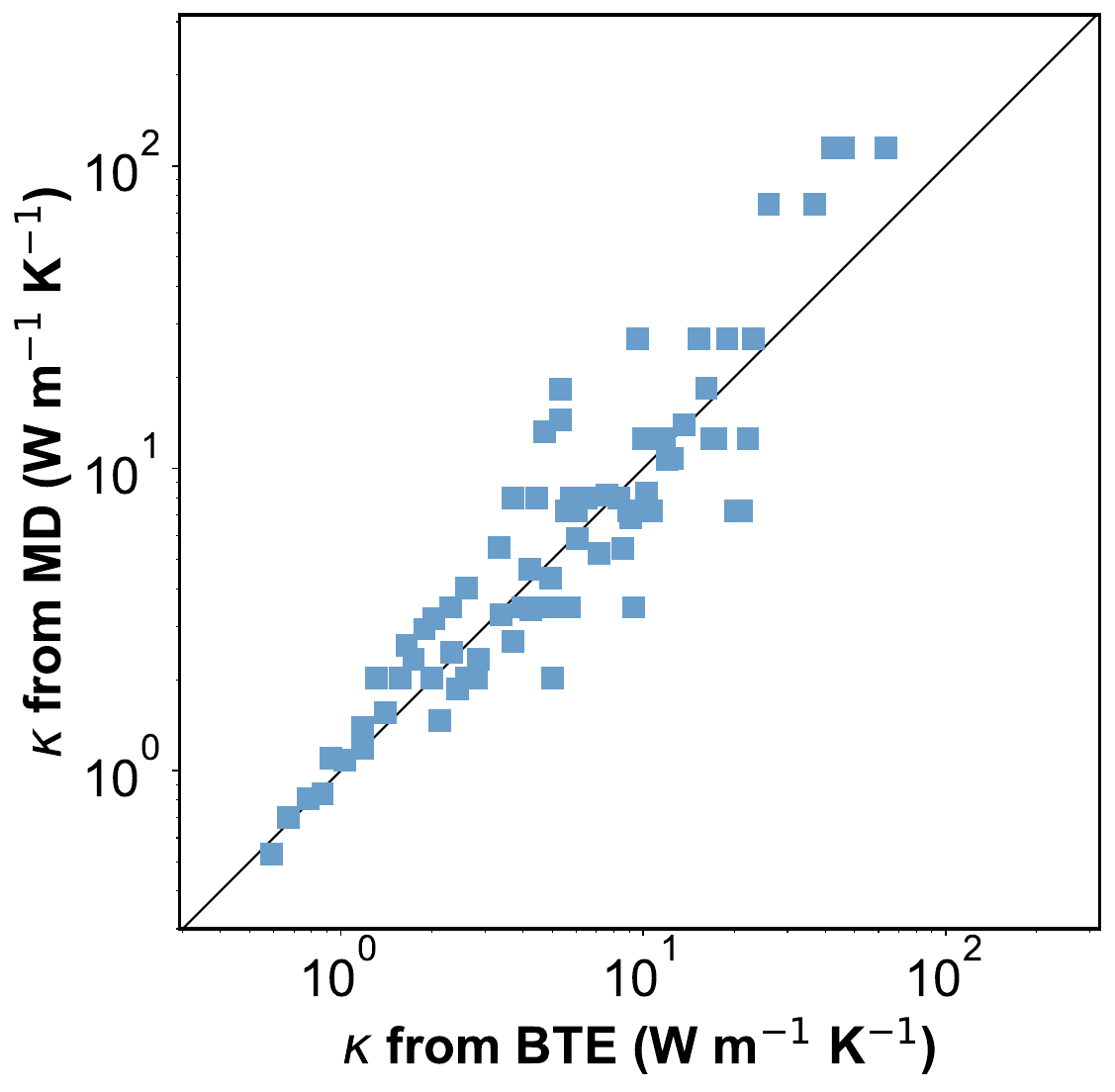}
\caption{Comparison of MD and BTE results for phonon thermal conductivities for all the materials considered in this work, including 16 metals and 4 binary alloys. Filled squares represent MD results from the UNEP-v1 model. Both phonon-phonon and phonon-electron scatterings are considered in the BTE results.}
\label{figure:parity-with-pe}
\end{center}
\end{figure}

\subsection{Comparison between NEP-MD and DFT-BTE results}

BTE results for phonon thermal conductivity are available for a number of metals across a broad temperature range, as shown in Fig.~\ref{figure:temperature}.
The BTE results for W and Mo are from Wang and Bao \cite{BaoHua2024APL}, while those for the remaining elements are from Wang \textit{et al.} \cite{WangYan2016JAP}.
For all these BTE calculations, isotope scattering has been considered. 
While phonon-electron scattering cannot be included into our current MD simulations, it can be considered in the BTE calculations. 
Therefore, for some metals, there are two sets of BTE results, one considering phonon-electron scattering and one without considering it. 
There are noticeable effects of phonon-electron scattering in Ni and Pt, but nearly no effects in Cu, Au, Al, and Ag \cite{WangYan2016JAP,Jain2016PRB}.
Focusing on the case without considering phonon-electron scattering, we see that the MD results based on the UNEP-v1 model are generally much closer to the BTE results than the MD results based on the EAM potential. 

Apart from elemental metals, we also considered a few binary alloys, including Cu$_{3}$Au, CuAu, Ni$_{3}$Al, and NiAl, which have been studied using the BTE approach \cite{}.
The results are listed in Table \ref{table:kappa}.
For these binary alloys, the MD results based on the UNEP-v1 model also show generally better agreement with the BTE results than the MD results based on the EAM potential.
To get an overall picture of the relative accuracy of the UNEP-v1 model compared to the EAM potential in the prediction of the phonon thermal conductivity, we plot all the MD results against available BTE results without considering phonon-electron scattering in Fig.~\ref{figure:parity}. 
The MD results based on the UNEP-v1 models show a clearly better correlation with the BTE results.
Taking the BTE results as references, the UNEP-v1 results have a root-mean-square error (RMSE) of 14.3 W m$^{-1}$ K$^{-1}$, while the EAM results exhibit a much larger RMSE of 20.3 W m$^{-1}$ K$^{-1}$.

Figure~\ref{figure:parity-with-pe} compares the HNEMD results based on the UNEP-v1 model and the BTE results considering both phonon-phonon and phonon-electron scatterings. 
The correlation between them is still reasonably good, which means that phonon-electron scattering does not dominate the determination of the phonon thermal conductivity in metals and alloys.
In view of this, we anticipate that the UNEP-v1 model is a promising tool in studying phonon thermal transport in more complex materials based on these metals, such as high-entropy alloys.

\section{Summary and Conclusions \label{section:summary}}

In summary, we have systematically calculated the phonon thermal conductivity of 16 metals and a few binary alloys using the HNEMD method, based on the recently developed unified neuroevolution potential for 16 metals and their alloys, UNEP-v1 \cite{song2024nc}.
Isotope disorder induced phonon scattering is found to be important in BCC metals such as Mo, W, and Cr.
We carefully compared our HNEMD simulation results against literature results based on the BTE approach considering phonon-phonon scattering only and found that the UNEP-v1 model significantly outperforms the empirical embedded-atom method potential. 
Our results establish the reliability of the UENP-v1 model in the calculation of phonon thermal transport properties in elemental metals and alloys.
Even though phonon-electron scattering is missing in our HNEMD simulations, our results show that this does not lead to strong deviations between our HNEMD results and the reference BTE results. 
Therefore, the UNEP-v1 model is a promising tool in studying phonon thermal transport in more complex materials such as high-entropy alloys, where phonon-defect or phonon-disorder scattering might play a strong role and phonon-electron scattering plays an even weaker role.

\vspace{0.5cm}
\noindent{\textbf{Data availability:}}

The data that support the findings of this study are available from the corresponding authors upon reasonable request.

\vspace{0.5cm}
\noindent{\textbf{Conflicts of interest:}}

The authors have no conflicts to disclose.

\begin{acknowledgments}
This work was supported by the National Science and Technology Advanced Materials Major Program of China (No. 2024ZD0606900). 
YS was supported by the Fundamental Research Funds for the Central Universities (FRF-TP-22-106A1).
HB was supported by the National Natural Science Foundation of China (Grant No. 52467080).
\end{acknowledgments}

\end{document}